\documentclass[twocolumn,showpacs,preprintnumbers,amsmath,amssymb]{revtex4} \usepackage{dcolumn}

\usepackage{graphicx} \usepackage{bm}

\newcommand{\hH}{\hat{H}}
\newcommand{\ha}{\hat{a}}
\newcommand{\hv}{\hat{v}}
\newcommand{\hp}{\hat{p}}
\newcommand{\hP}{\hat{P}}
\newcommand{\hx}{\hat{x}}
\newcommand{\psink}{\psi_{n\bm k}}
\newcommand{\psinkr}{\psi_{n\bm k}(\bm r)}
\newcommand{\unkr}{u_{n\bm k}(\bm r)}
\newcommand{\unkz}{u_{n 0}}
\newcommand{\unkrz}{u_{n 0}(\bm r)}
\newcommand{\unk}{u_{n\bm k}}
\newcommand{\chinkr}{\chi_{n\bm k}(\bm r)}

\newcommand{\epsnk}{\epsilon_n(\bm k)}

\newcommand{\gammar}{{\bm \gamma}({\bm r})}
\newcommand{\nablar}{{\bm \nabla}_{\bm r}}
\newcommand{\nablak}{{\bm \nabla}_{\bm k}}

\begin{document}
\title{Electron dynamics in crystalline semiconductors}
\date{\today}

\author{Wlodek Zawadzki}
\affiliation{Institute of Physics, Polish Academy of Sciences, Al. Lotnik\'ow 32/46,
             02-688 Warsaw, Poland}
\email{zawad@ifpan.edu.pl}

\begin{abstract}
Electron dynamics in crystalline semiconductors is described by distinguishing between an
instantaneous velocity related to electron's momentum
and an average velocity related to its quasi-momentum in a periodic potential.
It is shown that the electron velocity used in the theory of electron transport
 and free-carrier optics is the average electron velocity,
not the instantaneous velocity. An effective mass of charge carriers
in solids is considered and it is demonstrated that,
in contrast to the "acceleration" mass introduced in textbooks, it is a "velocity"
mass relating carrier velocity
to its quasi-momentum that is a much more useful physical quantity. Among other advantages,
the velocity mass is a scalar for spherical but nonparabolic energy bands~$\epsilon(k)$,
whereas the acceleration mass is not a scalar.
Important applications of the velocity mass are indicated.
A two-band~${\bm k}\cdot {\bm \hp}$ model is introduced
as the simplest example of a band
structure that still keeps track of the periodic lattice potential.
It is remarked that the two-band model, adequately describing narrow-gap
semiconductors (including zero-gap graphene),
strongly resembles the special theory of relativity.
Instructive examples of the "semi-relativistic" analogy are given.
The presentation has both scientific and pedagogical aspects.
\end{abstract}


\maketitle

\section{Introduction}

The phenomenon of Zitterbewegung (ZB, trembling motion), devised in 1930
by Erwin Schrodinger~\cite{Schroedinger1930},
has been for the last 80 years a subject of controversy and excitement.
The interest in this phenomenon experienced a strong revival in 2005,
when it was demonstrated that the trembling motion can occur also in
solids~\cite{Zawadzki2005,Schliemann2005}.
Since then there has been a real surge of papers proposing ZB in various periodic systems,
as reviewed in~\cite{Zawadzki2011}.
The nature of ZB in solids was investigated and it was shown that,
in its "classical" form analogous to ZB in a vacuum~\cite{Schroedinger1930},
the trembling motion represents oscillations of velocity when an electron
moves in a periodic potential of the lattice~\cite{Zawadzki2010}.
The situation resembles a roller-coaster: when the train moves upwards gaining the potential energy,
it slows down;
when the train goes down losing the potential energy, it accelerates.
However, in the solid state literature the largely prevailing picture is based on the Bloch theorem
in which electrons are treated as quasi-free particles with a modified (effective) mass.
This picture suggests that the electrons move in a solid with a constant velocity.
The same approach is used in the transport theory, in which carrier velocity is assumed constant and equal
to~${\bm v}=\hbar {\bm k}/m^*$, where~${\bm k}$ is the wave vector and~$m^*$ is the effective mass.
To the author's knowledge, there exist only two textbooks discussing an instantaneous
carrier velocity in a crystal: "Wave Mechanics of Crystalline Solids" by R. A. Smith~\cite{SmithBook}
and "Semiconductor Physics" by P. S. Kireev~\cite{KireevBook}.

The problem arises: how to reconcile the two pictures?
This is the first purpose of our work.
We show that the above question is related to the difference between carrier's momentum
and quasi-momentum in a periodic potential.
Our second purpose is to introduce properly an effective mass of carriers.
In solid state textbooks the effective mass is
always defined as a quantity relating an external force to carrier's acceleration.
We show that it is by far more useful to define the effective mass as a quantity
relating the average carrier velocity
to the quasi-momentum~$\hbar {\bm k}$. The "velocity mass" is a scalar for spherical
nonparabolic energy bands~$\epsilon(k)$,
whereas the "acceleration mass" is not. Important applications of the velocity mass are indicated.
In addition, we briefly describe a "semi-relativistic" behavior of charge carriers in
narrow-gap semiconductors including monolayer graphene.
This feature was discussed in the past but gained new significance
with "the rise of graphene"~\cite{Novoselov2004} and the advancement of Zitterbewegung.
To ensure the completeness and continuity of presentation,
we include in our exposition a few elements which are already
 known from the literature. As it stands, the text has both scientific and pedagogical aspects.

\section{Electrons In A Periodic Potential}

We begin by general considerations concerned with the motion of charge carriers in crystalline solids.
The Hamiltonian for an electron in a periodic potential~$V({\bm r})$ is
\begin{equation}
\hH= \frac{\hp^2}{2m_0} +V({\bm r}),
\end{equation}
where~$m_0$ is the free electron mass. The periodicity
signifies~$V({\bm r})=V({\bm r}+{\bm a})$ for~${\bm a}$ being a lattice vector.
The velocity operator is given by the Hamilton equation
\begin{equation}
\hv_i= \frac{\partial \hH}{\partial \hp_i} = \frac{\hp_i}{m_0}.
\end{equation}
The same result is obtained from the relation~$\hv_i=d\hx_i/dt = (1/i\hbar)[\hx_i,\hH]$.
The acceleration operator is
\begin{equation}
\ha_i= \frac{d \hv_i}{dt} = \frac{1}{i\hbar m_0}[\hp_i,\hH]
 = -\frac{1}{m_0} \frac{\partial V}{\partial \hx_i}
= \frac{1}{m_0}F_i^{pr},
\end{equation}
where~$F_i^{pr}=-\partial V/\partial \hx_i$ is a periodic force acting on
the electron moving in a periodic potential.
Equation~(3) is equivalent to the second Newton law of motion in an operator form.
It follows from Eq.~(3) that the momentum operator does not commute with the Hamiltonian~(1),
so it is not a constant of the motion.
In consequence, the velocity of Eq.~(2) and the acceleration of Eq.~(3)
are also not constants of the motion.
It is intuitively clear that, since the potential and the resulting force in Eq.~(3) are
periodic in~${\bm r}$, the acceleration
and the velocity will also be periodic functions of~${\bm r}$.
This result has an elementary classical interpretation.
Classically, the total electron energy is~$\epsilon=mv^2/2+V({\bm r})$,
so if the potential energy oscillates,
the kinetic energy (i.e., the velocity) also oscillates to keep the total energy constant.
This is, in fact, the physical origin of the trembling motion of electrons
in crystalline solids, see Ref.~\cite{Zawadzki2010}.
The results given in Eqs.~(2) and~(3) apply to any Hamiltonian with a scalar potential.
The specificity of a crystalline solid is that the potential is periodic,
so the Bloch theorem applies. Thus
\begin{equation}
\hH\psinkr = \epsnk\psinkr,
\end{equation}
where~$\epsnk$ is the energy of the~$n$-th band depending on the wave vector~${\bm k}$.
The Bloch state is~$\psinkr=\exp(i{\bm k}\cdot {\bm r})\unkr$ in
which the Bloch amplitude has the same periodicity as the potential in Eq.~(1),
i.e.~$\unkr=u_{n\bm k}(\bm r+\bm a)$. The quantity~$\hbar {\bm k}$ is an
eigenvalue of the quasi-momentum operator~${\bm \hP}$,
which should be carefully distinguished from the standard momentum
operator~${\bm \hp}$ introduced in Eq.~(1).
The Hamiltonian~(1) is invariant with respect to transformations possessing the symmetry of the potential
and there should exist a constant of the motion corresponding to this invariance.
This constant of the motion is precisely~$\hbar {\bm k}$. It means that the Bloch state~$\psinkr$
should also be an eigenstate of the quasi-momentum~${\bm  \hP}$, i.e. there should be by definition
\begin{equation}
{\bm \hP}\psinkr = \hbar {\bm k}\psinkr.
\end{equation}
We try to find an explicit expression for~${\bm \hP}$ looking
for the quasi-momentum operator in the form (see Ref.~\cite{KireevBook})
\begin{equation}
{\bm \hP}= {\bm \hp} + i\hbar \gammar,
\end{equation}
in which~$\gammar$ is a function of coordinates. We have
\begin{eqnarray}
{\bm \hP}\psink &=& \hbar{\bm k}\psink +
    i\hbar \gammar\psink-i\hbar e^{i{\bm k} \cdot {\bm r}} \nablar\unk \nonumber \\
         &=& \hbar{\bm k}\psink + i\hbar[\gammar - \nablar( \ln \unk)] \psink.
\end{eqnarray}
Putting~$\gammar=\nablar(\ln \unk)$ we get~${\bm \hP}\psink=\hbar{\bm k}\psink$,
so that Eq.~(5) is satisfied if
\begin{equation}
{\bm \hP}= -i\hbar{\bm \nabla}_{\bm r} + i\hbar{\bm \nabla}_{\bm r}(\ln\unk).
\end{equation}
In the second term in Eqs.~(7) and~(8) the differentiation acts only on the expression in parentheses.
Equation~(8) is instructive, as it shows explicitly that the operators of
momentum and quasi-momentum are distinctly different.
By using Eqs.~(4) and~(5) one easily shows that~$\hH{\bm \hP}={\bm \hP}\hH$,
so that the quasi-momentum~$\hbar{\bm k}$ is really a constant of the motion. To say it differently
\begin{equation}
\frac{d{\bm \hP}}{dt}=\frac{1}{i\hbar}[{\bm \hP},\hH]=0,
\end{equation}
which means that the periodic potential~$V({\bm r})$ in Eq.~(1)
does not change the quasi-momentum~${\bm \hP}$
whereas, as follows from Eq.~(3), it periodically changes the momentum~${\bm \hp}$
and velocity~${\bm \hv}$.
Still, the electron does not radiate because it is in the Bloch eigenenergy state.
Suppose now that, in addition to the periodic potential~$V({\bm r})$,
the electron experiences an additional
nonperiodic potential~$U^{ex}({\bm r})$.
This potential can be due to an external field, an impurity, a defect, etc.
Then the total Hamiltonian is
\begin{equation}
\hH_{tot}= \frac{\hp^2}{2m_0} +V({\bm r})+U^{ex}({\bm r}).
\end{equation}
It is easy to see that
\begin{eqnarray}
\frac{d\hp_i}{dt} &=& F_i^{pr} + F_i^{ex}, \\
\frac{d\hP_i}{dt} &=& F_i^{ex},
\end{eqnarray}
where~${\bm F}^{ex} = -\nablar U^{ex}({\bm r})$.
Thus the momentum is changed by both periodic and nonperiodic potentials,
whereas the quasi-momentum is changed only by the additional nonperiodic potential.

The question arises how to reconcile the oscillating electron velocity~${\bm \hv(t)}$
described in Eqs.~(2) and~(3) with the velocity appearing,
for example, in the transport phenomena and other kinetic effects in crystalline solids.
The time-dependent instantaneous velocity is given in general in the Heisenberg picture by
\begin{equation}
{\bm\hv}(t)= \exp(i\hH t/\hbar){\bm\hv}\exp(-i\hH t/\hbar),
\end{equation}
and it is this velocity operator that leads to the trembling motion, see Ref.~\cite{Zawadzki2011}.
Let us calculate an average of~${\bm\hv}(t)$ on the Bloch state~$\psinkr$.
We obtain by a simple manipulation
\begin{eqnarray}
\bar{\bm v}&=&\langle \psink|e^{i\hH t/\hbar}\hv e^{-i\hH t/\hbar}|\psink\rangle \nonumber \\
     &=&\langle \psink |e^{i\epsilon_{n{\bm k}} t/\hbar}{\bm\hv}
        e^{-i\epsilon_{n{\bm k}} t/\hbar}|\psink\rangle =
        \langle \psink|{\bm\hv}|\psink\rangle, \ \ \
\end{eqnarray}
where~${\bm \hv}={\bm \hp}/m_0$. Thus the average velocity in the Bloch state is time independent
because of the basic property of Eq.~(4).
The average velocity has been calculated in various ways, see Refs.~\cite{AshcroftBook,AnselmBook,KittelBook}.
Below we use the method based on the Hellmann-Feynman theorem~\cite{AslangulBook}.
Let us first write the Schrodinger equation
for the Bloch amplitude. As follows from Eq.~(4)
\begin{equation}
\hH_u({\bm r}, {\bm \hp};{\bm k})\unkr=\epsnk\unkr,
\end{equation}
where
\begin{equation}
\hH_u({\bm r}, {\bm \hp};{\bm k})=\frac{1}{2m_0}({\bm \hp}+\hbar {\bm k})^2+V(\bm r),
\end{equation}
depends parametrically on~${\bm k}$. We have
\begin{eqnarray}
\nablak \epsnk&=& \langle \unk|\nablak \hH_u|\unk\rangle =
             \langle \unk|\frac{\hbar}{m_0}({\bm \hp}+\hbar {\bm k})|\unk\rangle \nonumber \\
              &=& \frac{\hbar^2}{m_0} \langle \unk|-i\nablar +{\bm k}|\unk\rangle,
\end{eqnarray}
in which the first equality follows from the Hellmann-Feynman theorem. Further
\begin{eqnarray}
(-i\nablar +{\bm k})\unk &=&  (-i\nablar +{\bm k})e^{-i{\bm k} \cdot {\bm r}} \psink \nonumber \\
                         &=&  -ie^{-i{\bm k} \cdot {\bm r}} \nablar \psink,
\end{eqnarray}
so that
\begin{eqnarray}
\frac{1}{\hbar}\nablak \epsnk &=& -i\frac{\hbar}{m_0}\langle \unk|e^{-i{\bm k} \cdot {\bm r}}
             \nablar |\psink \rangle \nonumber \\
               &=& \langle\psink|\frac{{\bm \hp}}{m_0}|\psink\rangle = \bar{\bm v}_n(k).
\end{eqnarray}
The result~(19) is simple and important. It relates the average of instantaneous velocity in
the Bloch state~$\langle\psink|{\bm \hp}/m_0|\psink\rangle$ to
the electron energy~$\epsnk$ given as a function of the quasi-momentum~$\hbar{\bm k}$.
Below we will consider a specific energy band, so we drop the band index~$n$.

Now we want to associate the above results with an effective mass of charge carriers in an energy band.
In textbooks one considers standard parabolic and spherical energy bands described
by the energy-wave vector relation~$\epsilon=\hbar^2k^2/2m_0^*$,
where~$m_0^*$ is a constant effective mass at the band edge. It is then shown that
such a mass relates carrier's acceleration to an external force.
However, we want to consider a more general case of spherical but nonparabolic
energy bands in which the energy depends
on the absolute value of the wave vector in an arbitrary way, i.e.~$\epsilon=\epsilon(k)$.
In fact, many III-V semiconducting compounds (InSb, InAs, GaSb, GaAs, InP)
as well as II-VI compounds (HgTe, CdTe, HgCdTe, HgSe)
and their alloys possess the conduction bands of this type.
In contrast to the procedure adopted in textbooks, we define an effective mass
not by a relation between an external
force and acceleration, but as a quantity relating the average velocity~$\bar{\bm v}$
to the quasi-momentum~$\hbar{\bm k}$.
Thus we {\it define the effective mass} by the equality
\begin{equation}
m^*\bar{\bm v}=\hbar{\bm k},
\end{equation}
where~$\bar{\bm v}$ is given by Eq.~(14).
Since~$\bar{\bm v}$ and~$\hbar{\bm k}$ are vectors, the mass~$m^*$ is in principle a~$3\times 3$ tensor.
Using Eq.~(19) and the sphericity of the band we calculate
\begin{equation}
\bar{v}_i=\frac{\partial \epsilon}{\hbar \partial k_i}=
         \frac{d \epsilon}{\hbar dk}\frac{\partial k}{\partial k_i}=
          \frac{d\epsilon}{\hbar dk} \frac{k_i}{k}  =
          \frac{d\epsilon}{\hbar dk} \frac{1}{k}\delta_{ij}k_j,
\end{equation}
where in the last term we adopt the sum convention over the repeated coordinate subscript~$j=1,2,3$.
Using the definition~(20), the inverse mass tensor is
\begin{equation}
\bar{v}_i=\left(\frac{1}{m^*}\right)_{ij}\hbar k_j.
\end{equation}
By equating Eq.~(21) with Eq.~(22) we obtain
\begin{equation}
\left(\frac{1}{m^*}\right)_{ij}=\frac{d\epsilon}{\hbar^2 dk}\frac{1}{k}\delta_{ij}.
\end{equation}
Thus the inverse mass tensor is a {\it scalar} for a spherical energy band
\begin{equation}
\frac{1}{m^*}=\frac{1}{\hbar^2k}\frac{d\epsilon}{dk}.
\end{equation}
The average velocity is finally, see Eq.~(22),
\begin{equation}
\bar{\bm v}=\frac{\hbar{\bm k}}{m^*}.
\end{equation}
Recalling that~$\bar{\bm v}=\bar{\bm p}/m_0$, see Eq.~(19), we can write
\begin{equation}
\frac{\bar{\bm p}}{m_0}=\frac{\hbar{\bm k}}{m^*},
\end{equation}
which shows an analogy between the average momentum in the Bloch state~$\psi_{\bm k}$
and the quasi-momentum.
However, it is known that for electrons and light holes in semiconductors there is usually~$m_0\gg m^*$,
so that~$\bar{\bm p} \gg \hbar{\bm k}$.
This shows once again the difference between momentum and quasi-momentum.

Equation~(25) represents the basic formula for velocity used in the description of charge
carriers in semiconductors and metals.
Here we have obtained this formula with two important qualifications.
First, on the left-hand side we have the
{\it average velocity}
of a carrier in the Bloch state, not the instantaneous velocity
considered in the beginning, see Eq.~(2).
Second, on the right-hand side we have the velocity effective mass defined in Eq.~(20).
This mass depends in general on carrier's energy (or wave vector).
Since the average velocity~$\bar{\bm v}$ is expressed
by the {\it first derivative}~$d\epsilon/dk$, the velocity effective mass~$m^*$ is also related
to the first derivative~$d\epsilon/dk$.
On the other hand, the "acceleration" effective mass~$M_{ij}$ relating force to acceleration,
as introduced in textbooks,
is given by the second derivative of energy with respect to~${\bm k}$,
so this mass does not enter into the basic formula~(25),
unless one takes the simplest energy band described by~$\epsilon=\hbar^2k^2/2m_0^*$.
As is easy to see, in this particular case both masses are equal to~$m^*_0$.
We emphasize that the velocity mass, defined in Eq.~(20),
is much more useful than the acceleration mass defined in textbooks.
In particular, it is the velocity mass that defines carrier's
 mobility and is measured in various experiments.
We discuss this point below.

To conclude this section, we relate the first derivative~$d\epsilon/dk_i$ to the
group velocity of a carrier in a periodic potential.
Let us form a wave packet of Bloch states
\begin{eqnarray}
f(\bm r)&=&\int a({\bm k})e^{i{\bm k} \cdot {\bm r}} u_{\bm k}(\bm r) d^3{\bm k} \nonumber \\
        & \simeq & u_{\bm k_0}(\bm r) \int a({\bm k})e^{i{\bm k} \cdot {\bm r}} d^3{\bm k}.
\end{eqnarray}
It is assumed that the packet is narrow in~${\bm k}$ space
and it is centered around the value of~${\bm k_0}$.
This means that in the coordinate space the packet extends over several unit cells.
Thus the amplitudes~$a({\bm k})$ in Eq.~(27)
are non-vanishing only for small values of~${\bm q}={\bm k}-{\bm k}_0$.
In consequence, it is possible to take an average value of~$u_{\bm k_0}(\bm r)$
out of the integral sign. We further have
\begin{equation}
f(\bm r)= u_{\bm k_0}({\bm r})e^{i{\bm k}_0 \cdot {\bm r}}
  \int a({\bm k})e^{i{\bm q} \cdot {\bm r}} d^3{\bm q}.
\end{equation}
The integrand on the right-hand side of Eq.~(28) is formally identical to that of a free particle
and one may apply to this wave packet
the well known arguments determining the group velocity, which gives
\begin{equation}
v_i^{gr}= \left.\frac{\partial\omega}{\partial k_i}\right|_{\bm k_0} =
    \left.\frac{\partial\epsilon}{\hbar\partial k_i}\right|_{\bm k_0},
\end{equation}
where the derivative is taken at~${\bm k}={\bm k}_0$.
Thus the average velocity given in Eq.~(19)
is also the group velocity of the carrier.

\section{Two-Band Model. Semirelativity. Graphene.}

Now we consider an instructive and sufficiently general example of a band
structure in semiconductors in
order to illustrate consequences of the above formalism. As mentioned in relation to Eqs.~(1) and~(4),
the Bloch states are solutions to the eigenenergy equation with the Hamiltonian having a periodic
potential of the crystal lattice.
However, it is known that, for questions related to the band structure near a specific
point of the Brillouin zone or to problems
of carriers in external fields, it is more practical to work with the Luttinger-Kohn~(LK)
representation (see Refs.~\cite{Luttinger1955} and~\cite{Zak1966}).
The LK functions are~$\chinkr=e^{i{\bm k}\cdot {\bm r}}\unkrz$,
where~$\unkrz$ are the Bloch periodic amplitudes
taken at a fixed point~${\bm k}_0$ of the Brillouin zone.
We take for simplicity~${\bm k}_0=0$, i.e. the zone center.
It is clear that the LK amplitudes satisfy the eigenenergy equation
\begin{equation}
\left[\frac{\hp^2}{2m_0}+V({\bm r})\right]\unkz=\epsilon_{n0}\unkz,
\end{equation}
where~$\epsilon_{n0}$ is the energy of the~$n$-th band at~${\bm k}=0$.
One can show that the LK functions form a complete orthogonal set,
so one can represent a Bloch state as
\begin{equation}
\psi_{n'{\bm k}}({\bm r})= e^{i{\bm k}\cdot {\bm r}} \sum_n c_n^{n'}({\bm k}) \unkrz,
\end{equation}
in which~$c_n^{n'}({\bm k})$ are~${\bm k}$-dependent coefficients.
The summation is over all bands~$n$.
What follows is the standard procedure of transforming a differential
eigenvalue equation into an algebraic problem.
By inserting the form~(31) into initial Eq.~(1), using Eq.~(30), multiplying on the left by~$u_{n'0}$
and integrating over the unit cell, one obtains
\begin{equation}
\sum_n \left[(-\epsilon' + \epsilon_{n0})\delta_{n'n}+\frac{\hbar}{m_0}{\bm k}\cdot {\bm p}_{n'n} \right]
   c_n^{n'} = 0,
\end{equation}
for~$n'=1,2,3\ldots$. Here~$\epsilon'=\epsilon - \hbar^2k^2/2m_0$
and~${\bm p}_{n'n}=\langle \unkz|{\bm \hp}|\unkz\rangle$
are the interband matrix elements of momentum. Equation~(32) represents an infinite
set of equations for~$c_n^{n'}$ coefficients and the condition of non-trivial solutions
determines the energies~$\epsilon_{n'}(\bm k)$.
We now assume that an energy gap~$\epsilon_g$ between the conduction and valence bands is
much smaller than other gaps of interest,
so we can neglect the distant bands and keep in Eq.~(32) only the two close bands.
In addition, we neglect the free electron term~$\hbar^2k^2/2m_0$ in the energy
as it is small compared to the effective mass term, see below.
Taking the zero of energy in the middle of the gap,
so that~$\epsilon_{10}=+\epsilon_{g}/2$ and~$\epsilon_{20}=-\epsilon_{g}/2$,
the set~(32) is reduced to
\begin{equation}
 \left(\begin{array}{cc}
 +\epsilon_{g}/2 & {\bm \pi}_{12}\cdot \hbar {\bm k} \\
  {\bm \pi}_{21}\cdot \hbar {\bm k} & -\epsilon_{g}/2
 \end{array} \right) \left(\begin{array}{c} c_1 \\ c_2 \end{array} \right) =
 \epsilon \left(\begin{array}{c} c_1 \\ c_2 \end{array} \right),
\end{equation}
where~${\bm \pi}_{12}={\bm p}_{12}/m_0$ and similarly for~${\bm \pi}_{21}$.
Solving the above set for the energies one obtains
\begin{equation}
\epsilon(k) = \pm \left[\left(\frac{\epsilon_g}{2}\right)^2 +\epsilon_g
 \frac{\hbar^2k^2}{2m_0^*}\right]^{1/2},
\end{equation}
if we assume the simplest symmetry of the matrix elements
giving~$2{\bm \pi}_{12}{\bm \pi}_{12}/\epsilon_g=(1/m_0^*)\delta_{ij}$.
Here~$m_0^*$ is the electron effective mass at the band edge.
Plus and minus signs correspond to the conduction and valence bands, respectively.
Bands described by Eq.~(34) are spherical and nonparabolic.
For~$\hbar^2 k^2/2m_0^*\ll \epsilon_g/2$ one
can expand the square root and obtain~$\epsilon(k)=\pm (\epsilon_g/2 +\hbar^2 k^2/2m_0^*)$,
so that for small~$k$ values the bands are parabolic,
while for large~$k$ values they are linear in~$k$. Using for~$\epsilon(k)$
relation~(34) one can easily calculate the energy
dependence of the velocity mass~$m^*$ given by Eq.~(24). For the conduction band one obtains
\begin{equation}
m^*=m_0^*\frac{2\epsilon}{\epsilon_g}.
\end{equation}
At the band edge~$\epsilon=+\epsilon_g/2$ there is~$m^*=m_0^*$, as it should be.
The band-edge mass~$m_0^*$ in most semiconducting materials is much
smaller than the free electron mass~$m_0$,
so neglecting the free electron term~$\hbar^2k^2/2m_0$ in Eq.~(32) was justified.

It was remarked that the two-band~${\bm k} \cdot {\bm \hp}$ model~(2BM)
for the band structure of semiconductors closely resembles the
description of free relativistic electrons in a vacuum~\cite{Zawadzki1970,Zawadzki1997,Zawadzki2006}.
The Hamiltonian~(33), having the quasi-momentum terms off the diagonal,
looks very much like the Dirac equation without spin, while the dispersion~(34)
is analogous to the relativistic relation~$E=\pm \sqrt{(m_0c^2)^2+c^2p^2}$
with the correspondence~${\bm \hp} \rightarrow \hbar{\bm k}$ and
\begin{equation}
2 m_0c^2 \rightarrow \epsilon_g \hspace*{2em} m_0 \rightarrow m_0^*.
\end{equation}
It is easy to determine the maximum velocity~$u$ in the 2BM
\begin{equation}
c=\left(\frac{2m_0c^2}{2m_0} \right)^{1/2} \rightarrow
 \left(\frac{\epsilon_g}{2m_0^*} \right)^{1/2} = u.
\end{equation}
In light of our previous considerations,~$u$ is the maximum average velocity in the Bloch state.
The value of~$u$ can be determined by measuring the energy gap~$\epsilon_g$
and the band-edge mass~$m_0^*$ in a semiconductor material.
It turns out that the velocity~$u$ is almost the same in different materials
and is given by~$u\approx 10^8$ cm/s, i.e. it
is about 300 times smaller than the maximum velocity for relativistic electrons in a vacuum~$c$.
Using Eq.~(37) for~$u$ one can rewrite Eq.~(35) in the form
\begin{equation}
\epsilon = m^* u^2.
\end{equation}
This is equivalent to the famous Einstein formula:~$E=mc^2$ relating the energy to the mass.
The Compton wavelength~$\lambda_c=\hbar/m_0c$, playing an important role in the relativistic
quantum mechanics, also has a corresponding length
in the two-band~${\bm k} \cdot {\bm \hp}$ model, see Ref.~\cite{Zawadzki2005}
\begin{equation}
\lambda_Z=\frac{\hbar}{m_0^*u}=\hbar\left( \frac{2}{m_0^*\epsilon_g} \right)^{1/2}.
\end{equation}
This length determines the amplitude of Zitterbewegung oscillations mentioned
in the Introduction, see~\cite{Zawadzki2005}. In narrow-gap semiconductors
one can have~$m_0^*=5\times 10^{-2} m_0$ and, since~$c\approx 300u$,
one obtains~$\lambda_Z\approx 2\times 10^4 \lambda_c \approx 50$\AA, i.e.
a sizable length for nanostructures.
The dispersion relation~(34) can be rewritten in terms of~$u$ and~$\lambda_Z$ in the form
\begin{equation}
\epsilon(k)=\pm\hbar u \left(\lambda^{-2}_Z + k^2\right)^{1/2}\;\;.
\end{equation}

\begin{figure}
\includegraphics[width=8.5cm,height=8.5cm]{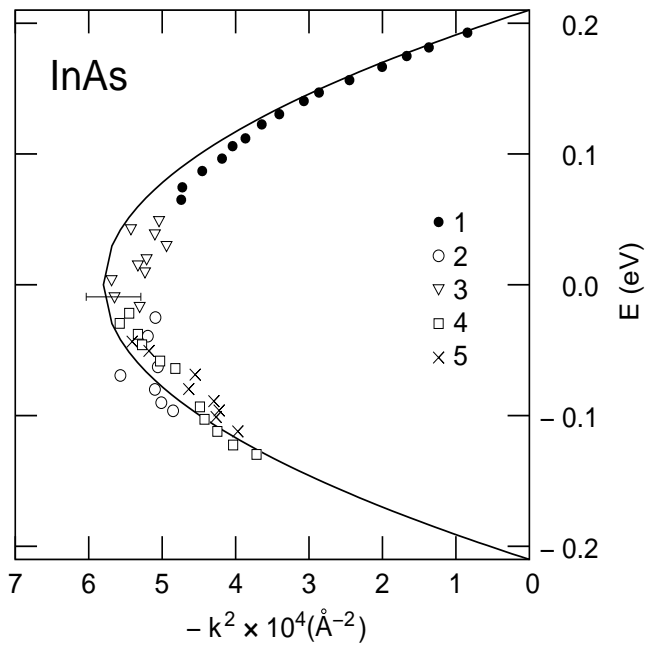}
\caption{\label{ZawadzkiFig1}Energy-wave vector dependence in the
forbidden gap of InAs. Various symbols show experimental data of
Parker and Mead~\cite{Parker1968}, the solid line is
theoretical fit using Eq.~(40). The determined
parameters are~$\lambda_Z$ = 41.5 {\rm \AA} and~$u$ = 1.33$\times 10^8$ cm/s.
After Ref.~\cite{Zawadzki2005}.}
\end{figure}
For~$k^2\ge 0$, Eq.~(40) describes the conduction and light-hole bands.
For~$k^2< 0$, that is for imaginary values of~$k=i\kappa$, this equation describes the
dispersion in the gap. The latter can be determined in metal-semiconductor
tunnelling experiments. In Fig.~1 we show the results of Parker and Mead~\cite{Parker1968} for
InAs, as described by Eq.~(40) with the use of adjustable parameters~$u$ and~$\lambda_Z$.
One obtains a very good description which confirms the validity of the two-band
model for narrow-gap materials. In particular, one obtains the above mentioned
large value of~$\lambda_Z$.

Regarding the phenomenon of Zitterbewegung we want to emphasize the following subtle point.
If one calculates the velocity operator using the matrix
Hamiltonian~(33):~$\hv_i=\partial \hH/\partial \hbar k_i$,
the velocity matrix does not commute
with the Hamiltonian:~$\hv_i\hH - \hH\hv_i \neq 0$, so the velocity depends on time
and it is the {\it instantaneous velocity}
containing the Zitterbewegung, see Refs.~\cite{Zawadzki2011,Rusin2007}.
However, if one calculates the velocity using
the energy~(34):~$\bar{v}_i=\partial \epsilon/\partial \hbar k_i$,
it is the {\it average velocity} not depending on time and given by Eq.~(25).
This means that the two-band model in the matrix form still "keeps track" of
the periodic Hamiltonian~(1) from which it originates, because both give the
Zitterbewegung. On the other hand, in the energy~(34) the track of periodicity of the original
Hamiltonian~(1) is already lost. One can use the LK transformation
to separate the conduction and valence
bands in the Hamiltonian~(33), which gives~$\epsilon_{\pm}=\pm (\epsilon_g/2 + \hbar^2k^2/2m_0^*)$
corresponding to the above mentioned expansion of the square root in Eq.~(34).
In this case the velocity
is~$\bar{v}_i^{\pm}=\partial \epsilon_{\pm}/\partial \hbar k_i= \pm\hbar k_i/m_0^*$,
i.e. it is the average velocity for each band with no Zitterbewegung.
Thus the two-band model is the simplest~${\bm k}\cdot {\bm \hp}$ description
reproducing the essential features of the initial periodic Hamiltonian~(1).

Finally, we want to consider briefly the important case of monolayer two-dimensional
graphene in light of the above discussion.
Graphene's band structure near the K point of the Brillouin zone is described
by the Hamiltonian~\cite{Slonczewski1958}
\begin{equation}
\hH = \hbar u \left(\begin{array}{cc} 0 & k_x - ik_y \\ k_x + ik_y & 0 \end{array}\right),
\end{equation}
where~$u\approx 10^8$ cm/s. The above form can be considered to be a special
case of the two-band model~(33) with the vanishing gap~$\epsilon_g=0$
and properly chosen matrix elements~$\pi_{12}^x$ and~$\pi_{12}^y$.
The resulting energy dispersion is linear in quasi-momentum:~$\epsilon=\pm u\hbar k$,
where~$k=\sqrt{k_x^2+k_y^2}$.
In view of our semi-relativistic analogy this case can be considered to be
the "extreme relativistic limit".
The matrix velocity operator~$\hv_i=\partial \hH/\partial \hbar k_i$ does not
commute with the Hamiltonian~(41)
and the instantaneous velocity contains the ZB component~\cite{Rusin2007}.
The velocity~$\bar{v_i}=\partial \epsilon/\partial \hbar k_i = u k_i/k$
represents an average velocity calculated in Eq.~(21).
The absolute value of velocity vector for any direction is~$\bar{v}=\sqrt{v_x^2+v_y^2}=u$.
The velocity mass can still be defined as before:~$1/m^*=(1/\hbar^2k)d\epsilon/dk$.
For the linear band dispersion one has~$d\epsilon/dk=u$, so that~$m^*=\hbar k/u=\epsilon/u^2$.
This gives, as before,~$\epsilon=m^*u^2$, see Eq.~(38).
One can also write~$m^*=\epsilon/u^2$ which means that at the band edge
(called in the literature "the Dirac point")
the effective mass is zero, but as the energy increases the mass increases as well.
Now let us suppose that an external force is applied along the~$x$ direction to
an electron characterized by~$k_y=0$.
According to Eq.~(12) there is~$d(\hbar k_x)/dt=F_x^{ex}$.
Since for~$k_y=0$ there is~$\hbar k_x=m^*\bar{v}_x=m^*u$. Thus the change of~$\hbar k_x$
due to the external force goes entirely into the change of the mass.

\section{Velocity And Acceleration Effective Masses}

In this section we consider the use of the velocity effective mass in spherical
and spheroidal energy bands. At the end we
mention some properties of the acceleration effective mass of charge carriers.
We begin with the velocity mass~$m^*$ which, as mentioned above,
is much more useful than the acceleration mass.
In our considerations below we are concerned with the average electron
motion related to the quasi-momentum,
so we drop the sign of "average" over the velocity, i.e. we write~$\bar{\bm v}= {\bm v}$.
The important property of the velocity mass is that it is measured in the cyclotron resonance~(CR).
We first demonstrate it using the classical electron motion in a magnetic field.
The equation of motion is
\begin{equation}
\frac{d(\hbar {\bm k})}{dt} = e({\bm v} \times {\bm B}),
\end{equation}
where~${\bm B}=[0,0,B]$ is a magnetic field applied along the~$z$ direction.
Using the definition~(20) of the velocity mass
for a spherical band:~$\hbar {\bm k}=m^*{\bm v}$, we have
\begin{equation}
m^* \frac{d{\bm v}}{dt} = e({\bm v} \times {\bm B}).
\end{equation}
Since, as we showed above,~$m^*$ for a nonparabolic band depends in general on electron energy,
one can imagine that~$m^*$ depends
also on time if the energy during the motion as not constant.
However, it is well known that a constant and
uniform magnetic field does not do any work, so the electron energy is a constant.
For this reason we assumed~$m^*$ not to depend on time in arriving at Eq.~(43).
For the first two components
the above equation gives
\begin{eqnarray}
\frac{dv_x}{dt}&=& \ \ \frac{eB}{m^*}v_y, \\
\frac{dv_y}{dt}&=& -\frac{eB}{m^*}v_x.
\end{eqnarray}
One can now differentiate Eq.~(45) with respect to time, insert the result into Eq.~(44)
and arrive at the second-order differential equation for~$v_y$,
which can be easily solved in terms of trigonometric functions.
Instead, we simply guess the solutions (see Ref.~\cite{KittelBook2}):~$v_x=v_0\cos(\omega t)$
and~$v_y=-v_0\sin(\omega t)$,
in which~$\omega$ is the cyclotron frequency with which the electron circles on the orbit.
Using the above solutions one obtains from Eqs.~(44) and~(45) the same result
\begin{equation}
\omega = \frac{eB}{m^*}.
\end{equation}
Thus the cyclotron frequency is determined by the velocity mass~$m^*$.
The cyclotron orbit can be obtained by integrating the velocity over time,
which gives:~$(x-x_0)^2+(y-y_0)^2=v_0^2/\omega^2$.
This means that the classical cyclotron radius is also determined by~$m^*$.
The same result concerning the velocity mass can be obtained
from the quantization of motion in a magnetic field.
To be specific, we use the orbital and spin quantization resulting from the band
structure of InSb-type III-V semiconducting compounds.
The band structure includes three levels at the~$\Gamma$ point of
the Brillouin zone (eight bands including spin).
The resulting quantized orbital and spin levels are~\cite{Bowers1959,Zawadzki1980}
\begin{equation}
\epsilon(n,k_z,\pm)=\left[\left(\frac{\epsilon_g}{2}\right)^2 +\epsilon_g D_{nk_z\pm}\right]^{1/2},
\end{equation}
where
\begin{equation}
D_{nk_z\pm} = \hbar \omega_c^0\left(n+\frac{1}{2}\right) +
 \frac{\hbar^2k_z^2}{2m_0^*} \pm \frac{1}{2} g_0^*\mu_B B.
\end{equation}
Here~$\omega_c^0=eB/m_0^*$, in which~$m_0^*$ is the band-edge mass [see Eq.~(34)],~$g_0^*$ is the
band-edge spin Lande factor and~$\mu_B$ is the Bohr magneton.
The cyclotron energy~$\hbar\omega$ is given by the energy difference between two consecutive
orbital levels:~$\epsilon(n+1,k_z,\pm) - \epsilon(n,k_z,\pm)$. Using Eqs.~(47) and~(48) we have
\begin{equation}
\hbar \omega = \frac{\epsilon(n+1)^2-\epsilon(n)^2}{\epsilon(n+1)+\epsilon(n)}
 = \frac{\hbar\epsilon_g eB}{m_0^*}\frac{1}{\epsilon(n+1)+\epsilon(n)}.
\end{equation}
For small magnetic fields there is~$\epsilon(n+1)+\epsilon(n) \approx 2\epsilon$,
so that~$\hbar \omega \approx \hbar eB/(2m_0^*\epsilon/\epsilon_g)= \hbar eB/m^*$, see Eq.~(35).
Thus, again, the cyclotron frequency is determined by the velocity effective mass~$m^*$.
The same reasoning can be applied to the so called "inverted " band structure of zero-gap
and narrow-gap II-VI compounds based on HgTe and HgSe, see~\cite{Zawadzki1980}.

As mentioned above, the basic relation for the classical transport
theory is~$\hbar{\bm k}=m^*{\bm v}$.
This leads to the definition of carrier's mobility~$\mu=q\tau/m^*$,
which involves the relaxation time~$\tau$
and the velocity mass~$m^*$~\cite{Zawadzki1974}. As a consequence, the mobility is directly
affected by the energy variation of~$m^*(\epsilon)$.
Some d.c. transport phenomena at high magnetic fields do
not depend on the relaxation time so that, by studying them,
one gains a direct access to the mass~$m^*$~\cite{Zawadzki1974}.
Finally, the free-carrier optics depends on the band structure
only through the velocity mass. And so the reflectivity depends on~$\langle 1/m^*\rangle$,
the magneto-reflectivity on~$\langle 1/m^{*2}\rangle/\langle1/m^*\rangle$,
the Faraday rotation is proportional to~$\langle 1/m^{*2}\rangle$ and
the Voigt phase shift to~$\langle 1/m^{*3}\rangle$.
Here the brackets denote appropriate averages over electron energies in the band~\cite{Zawadzki1974}.
The knowledge of~$m^*(\epsilon)$ gives then a direct information on the band structure.

Another strong indication, that the velocity effective mass is much more useful
than the acceleration mass, is the fact that the corresponding mass
is commonly used in the special theory of relativity~(STR).
In STR this mass is defined by the relation:~${\bm p}=m(v){\bm v}$,
it is a scalar and it has the famous velocity
dependence:~$m(v)=m_0/(1-v^2/c^2)^{1/2}$. Its energy dependence is not written down so often,
but it is not difficult to derive.
Since in STR the energy is given by~$E=[(m_0c^2)^2+p^2c^2]^{1/2}$,
the velocity is~$v_i=\partial E/\partial p_i = p_ic^2/E$,
and the velocity mass is~$m=E/c^2=2m_0E/(2m_0c^2)$.
It is seen that, using the semi-relativistic analogy:~$m_0 \rightarrow m_0^*$
and~$2m_0c^2 \rightarrow \epsilon_g$,
the relativistic velocity mass has the same energy dependence
as the effective velocity mass resulting from
the two-band~${\bm k} \cdot {\bm \hp}$ model, Eq.~(35).
We mention that the relativistic velocity-dependent mass~$m(v)$ is somewhat
reluctantly used in STR by some authors because of its
unorthodox transformation properties, see~\cite{UgarovBook}.
However, the transformation problem is not relevant for solids.

To conclude our considerations of the velocity mass we treat
an important case of ellipsoidal energy bands
which occur in semiconducting II-VI lead salts PbTe, PbSe, PbS,
as well as in silicon and germanium.
Such an energy band with arbitrary nonparabolicity
can be described by the relation~\cite{Zawadzki1974,Zukotynski1963}
\begin{equation}
\gamma(\epsilon)= a_{\alpha\beta} k_{\alpha}k_{\beta},
\end{equation}
where~$\gamma(\epsilon)$ is a "reasonable" function of energy
describing the nonparabolicity of the band.
The limiting assumption is that the shape of the ellipsoid
does not vary with the energy.
We use the sum convention over the repeated coordinate indices.
The tensor~$a_{\alpha\beta}$ is symmetric and it can be brought to a diagonal form by an
appropriate rotation of coordinates in~${\bm k}$ space.
Then the unequal diagonal components~$a_{ii}$
express band's ellipsoidal shape. An inverse tensor of velocity mass is
defined by the relation~(22). On the other hand,
since the velocity is~$v_i=\partial \epsilon/\partial \hbar k_i$,
one obtains with the help of Eq.~(50)
\begin{equation}
\left(\frac{1}{m^*}\right)_{ij}= \frac{2}{\hbar^2}
 \left( \frac{d\gamma}{d\epsilon}\right)^{-1}a_{ij}.
\end{equation}
For a parabolic band there is~$\gamma(\epsilon)\equiv \epsilon$,
and the inverse mass components are numbers.
A spherical band with arbitrary nonparabolicity is described
by~$a_{\alpha\beta}=\delta_{\alpha\beta}$,
the Kronecker delta.
This gives~$\gamma(\epsilon)=k^2$, [see Eq.~(50)],
which is equivalent to the spherical case considered above.

Finally, we calculate the inverse tensor of acceleration mass for
a spherical nonparabolic band~$\epsilon(k)$.
The general expression of the inverse mass tensor relating force
to acceleration is well known
\begin{equation}
\left(\frac{1}{M^*}\right)_{ij}= \frac{1}{\hbar^2}
  \frac{\partial^2 \epsilon}{\partial k_i\partial k_j}.
\end{equation}
A simple manipulation gives
\begin{equation}
\left(\frac{1}{M^*}\right)_{ij}= \frac{1}{\hbar^2 k}
   \frac{\partial \epsilon}{\partial k} \delta_{ij}+
\frac{k_i k_j}{k^2} \left( \frac{d^2\epsilon}{\hbar^2 dk^2} - \frac{1}{h^2 k}
  \frac{d\epsilon}{dk} \right).
\end{equation}
It is seen that, in contrast to the velocity mass, the acceleration mass is not a scalar
quantity even for a spherical energy band.
Interestingly, it is the band nonparabolicity that makes the acceleration mass non-scalar.
For a standard parabolic band:~$\epsilon=\hbar^2k^2/2m_0^*$,
the second term in Eq.~(53) vanishes,
so that~$(1/M^*)_{ij}=(1/m_0^*)\delta_{ij}$. This is identical with
the velocity mass for a parabolic band.
Only in this simple case the two masses coincide.
Because of the non-scalar character of the acceleration mass~(53)
the acceleration in a nonparabolic band is not parallel to the force.
This feature is well known in the special relativity,
which illustrates once again the semi-relativistic analogy.

\section{Conclusions And Summary}
We summarize our work by enumerating the main conclusions and
indicating the corresponding equations. For a carrier moving in a periodic potential,
the momentum, velocity, and acceleration are not constants of the motion,
see Eqs.~(1)-(3). The quasi-momentum, which is a distinctly different
operator from the momentum, see Eq.~(8),
is a constant of the motion in a Bloch state, see Eq.~(9).
The average electron velocity in a Bloch state is given by a
gradient of the energy with respect to the quasi-momentum~$\hbar {\bm k}$, see Eq.~(19).
It is this average velocity~$\bar{\bm v}$ which is used in the classical
transport theory for charge carriers.
A "velocity effective mass" is defined as a quantity
relating the average velocity to the quasi-momentum, see Eqs.~(20) and~(21).
The velocity mass for a spherical energy band is a scalar,
see Eq.~(23), and it enters into the basic relation
for the transport theory, see Eq.~(25).
A two-band~${\bm k}\cdot {\bm \hp}$ model in the matrix form
is the simplest description of the band structure
that still keeps track of the periodic potential, see Eq.~(33).
The two-band~${\bm k}\cdot {\bm \hp}$
Hamiltonian~(33) and the resulting energy~(34)
bear strong similarity to the description of free relativistic electrons in a vacuum.
In particular, they lead to an analog of the famous Einstein relation between
the mass and the energy, see Eq.~(38). In this perspective, the band structure of
gapless graphene can be regarded as an extreme relativistic case.
The velocity effective mass is much more useful than the acceleration mass commonly
introduced in solid state textbooks.
In particular, it is the velocity mass that is measured in the cyclotron resonance,
see Eqs.~(46) and~(49), in d.c. transport phenomena and in the free-carrier optics.
The velocity mass can also be introduced for ellipsoidal
nonparabolic energy bands, see Eq.~(51).

\acknowledgements
It is my pleasure to thank to Dr T.M. Rusin for elucidating discussions.


\begin{thebibliography}{99}
\bibitem{Schroedinger1930} E. Schrodinger, Sitzungsber. Preuss. Akad. Wiss. Phys.
                           Math. Kl. {\bf 24} 418, (1930).
                           Schrodinger's derivation is reproduced in A. O. Barut and A. J. Bracken,
                           Phys. Rev. D {\bf 23}, 2454 (1981).
\bibitem{Zawadzki2005}     W. Zawadzki, Phys. Rev. B {\bf 72}, 085217 (2005).
\bibitem{Schliemann2005}   J. Schliemann, D. Loss and R. M. Westervelt,
                           Phys. Rev. Lett. {\bf 94}, 206801 (2005).
\bibitem{Zawadzki2011}     W. Zawadzki and T. M. Rusin, J. Phys. Cond. Matt. {\bf 23}, 143201 (2011).
\bibitem{Zawadzki2010}     W. Zawadzki and T. M. Rusin, Phys. Lett. A {\bf 374}, 3533 (2010).
\bibitem{SmithBook}        R. A. Smith, {\it Wave Mechanics of Crystalline Solids}
                           (Chapman and Hall, London, 1961).
\bibitem{KireevBook}       P. S. Kireev, {\it Semiconductor Physics} (MIR Publishers, Moscow, 1975).
\bibitem{Novoselov2004}    K. S. Novoselov, A. K. Geim, S. V. Morozov, D. Jiang, Y. Zhang,
                           S. V. Dubonos, I. V. Grigorieva, and A. A. Firsov,
                           Science {\bf 306}, 666 (2004).
\bibitem{AshcroftBook}     N. W. Ashcroft and N. M. Mermin, {\it Solid State Physics}
                           (Holt, Rinehart and Winston, New York, 1976).
\bibitem{AnselmBook}       A. I. Anselm, {\it Introduction to the Theory of Semiconductors}
                           (Prentice Hall, Englewood, 1982).
\bibitem{KittelBook}       C. Kittel, {\it Quantum Theory of Solids} (Wiley, New York, 1963).
\bibitem{AslangulBook}     C. Aslangul, {\it Mecanique Quantique, Vol. 2} (De Boeck, Bruxelles, 2008).
                           In French.
\bibitem{Luttinger1955}    J. M. Luttinger and W. Kohn, Phys. Rev. {\bf 97}, 869 (1955).
\bibitem{Zak1966}          J. Zak and W. Zawadzki, Phys. Rev. {\bf 145}, 536 (1966).
\bibitem{Zawadzki1970}     W. Zawadzki, in {\it Optical Properties of Solids},
                           edited by E. D. Heidemenakis (Gordon and Breach, New York, 1970), p~179.
\bibitem{Zawadzki1997}     W. Zawadzki, in {\it High Magnetic Fields in the Physics of Semiconductors II},
                           edited by G. Landwehr and W. Ossau (World Scientific, Singapore, 1997), p~755.
\bibitem{Zawadzki2006}     W. Zawadzki, Phys. Rev. B {\bf 74}, 205439 (2006).
\bibitem{Parker1968}       G. M. Parker and C. A. Mead, Phys. Rev. Lett {\bf 21}, 605 (1968).
\bibitem{Rusin2007}        T. M. Rusin and W. Zawadzki, Phys. Rev. B {\bf 76}, 195439 (2007).
\bibitem{Slonczewski1958}  J. C. Slonczewski and P. R. Weiss, Phys. Rev. {\bf 109}, 272 (1958).
\bibitem{KittelBook2}      C. Kittel, W. D. Knight and M. A. Ruderman, {\it Mechanics}
                           (McGraw-Hill, New York, 1962).
\bibitem{Bowers1959}       R. Bowers and Y. Yafet, Phys. Rev. {\bf 115}, 1165 (1959).
\bibitem{Zawadzki1980}     W. Zawadzki, in {\it Narrow Gap Semiconductors. Physics and Applications},
                           edited by W. Zawadzki (Springer, Berlin, 1980), p.~85.
\bibitem{Zawadzki1974}     W. Zawadzki, Adv. in Physics {\bf 23}, 435 (1974).
\bibitem{Zukotynski1963}   S. Zukotynski and J. Kolodziejczak, Phys. St. Solidi (b) {\bf 3}, 990 (1963).
\bibitem{UgarovBook}       V. A. Ugarov, {\it Special Theory of Relativity} (MIR Publishers, Moscow, 1977).
                           Supplement by V. L. Ginzburg.
\end{thebibliography}
\end{document}